\documentclass[twocolumn, preprintnumbers,prb,aps,amssymb,showpacs,superscriptaddress]{revtex4}
\usepackage{graphicx}
\usepackage{dcolumn}
\usepackage{bm} \usepackage{color}
\begin{document}

\newcommand{\ie}{{\it i.e.}}
\newcommand{\eg}{{\it e.g.}}
\newcommand{\etal}{{\it et al.}}
\newcommand{\K}{Ba$_{1-x}$K$_x$Fe$_2$As$_2$}
\newcommand{\Co}{Ba(Fe$_{1-x}$Co$_x$)$_2$As$_2$}
\newcommand{\Isovalent}{Ba$_2$Fe$_2$(As,Ru$_{1-x}$Co$_x$)$_2$}

\newcommand{\Kzero}{$\kappa_0/T$}
\newcommand{\KN}{$\kappa_{\rm N}$}

\newcommand{\Tc}{$T_{\rm c}$}
\newcommand{\Hc}{$H_{\rm c2}$}

\newcommand{\units}{$\mu \text{W}/\text{K}^2\text{cm}$}
\newcommand{\p}[1]{\left( #1 \right)}


\title{Doping evolution of  the superconducting gap structure in the underdoped iron arsenide \K\ revealed by thermal conductivity}


\author{J.-Ph.~Reid}
\affiliation{D\'epartement de physique \& RQMP, Universit\'e de Sherbrooke, Sherbrooke, Qu\'ebec J1K 2R1, \textcolor{black}{Canada}}

\author{M.~A.~Tanatar}
\affiliation{Ames Laboratory, Ames, Iowa 50011, USA}

\author{X.~G.~Luo}
\affiliation{D\'epartement de physique \& RQMP, Universit\'e de Sherbrooke, Sherbrooke, Qu\'ebec J1K 2R1, \textcolor{black}{Canada}}

\author{H.~Shakeripour}
\affiliation{D\'epartement de physique \& RQMP, Universit\'e de Sherbrooke, Sherbrooke, Qu\'ebec J1K 2R1, \textcolor{black}{Canada}}
\affiliation{Department of Physics, Isfahan University of Technology, Isfahan 84156-83111, Iran} 

\author{S.~Ren\'e de Cotret} 
\affiliation{D\'epartement de physique \& RQMP, Universit\'e de Sherbrooke, Sherbrooke, Qu\'ebec J1K 2R1, \textcolor{black}{Canada}}

\author{A.~Juneau-Fecteau} 
\affiliation{D\'epartement de physique \& RQMP, Universit\'e de Sherbrooke, Sherbrooke, Qu\'ebec J1K 2R1, \textcolor{black}{Canada}}

\author{J.~Chang} 
\affiliation{D\'epartement de physique \& RQMP, Universit\'e de Sherbrooke, Sherbrooke, Qu\'ebec J1K 2R1, \textcolor{black}{Canada}}

\author{B.~Shen}
\affiliation{\textcolor{black}{Center for Superconducting Physics and Materials, National Laboratory of Solid State Microstructures
and Department of Physics, Nanjing University, Nanjing 210093, China}}

\author{H.-H.~Wen}
\affiliation{\textcolor{black}{Center for Superconducting Physics and Materials, National Laboratory of Solid State Microstructures
and Department of Physics, Nanjing University, Nanjing 210093, China}}
\affiliation{Canadian Institute for Advanced Research, Toronto, Ontario M5G 1Z8, \textcolor{black}{Canada}}

\author{H.~Kim}
\affiliation{Ames Laboratory, Ames, Iowa 50011, USA} 
\affiliation{Department of Physics and Astronomy, Iowa State University, Ames, Iowa 50011, USA }

\author{R.~Prozorov}
\affiliation{Ames Laboratory, Ames, Iowa 50011, USA} 
\affiliation{Department of Physics and Astronomy, Iowa State University, Ames, Iowa 50011, USA }

\author{N.~Doiron-Leyraud} 
\affiliation{D\'epartement de physique \& RQMP, Universit\'e de Sherbrooke, Sherbrooke, Qu\'ebec J1K 2R1, \textcolor{black}{Canada}}

\author{Louis Taillefer}
\altaffiliation{\textcolor{black}{E-mail: louis.taillefer@usherbrooke.ca}}
\affiliation{D\'epartement de physique \& RQMP, Universit\'e de Sherbrooke, Sherbrooke, Qu\'ebec J1K 2R1, \textcolor{black}{Canada}}
\affiliation{Canadian Institute for Advanced Research, Toronto, Ontario M5G 1Z8, \textcolor{black}{Canada}}

\date{\today}


\begin{abstract}

The thermal conductivity $\kappa$ of the iron-arsenide superconductor \K~was measured for heat currents parallel and perpendicular 
to the tetragonal \textcolor{black}{$c$ axis} 
\textcolor{black}{at temperatures} down to 50~mK 
\textcolor{black}{and}
in magnetic fields up to 15~T. 
Measurements were performed \textcolor{black}{on} samples with compositions 
\textcolor{black}{ranging from} optimal doping 
\textcolor{black}{($x=0.34$; \Tc~$=39$~K)}
down to \textcolor{black}{dopings deep} into the region \textcolor{black}{where antiferromagnetic order coexists with} superconductivity 
\textcolor{black}{($x=0.16$; \Tc~$=7$~K)}.
\textcolor{black}{In zero field, there is no} residual linear term in $\kappa(T)$ as $T\to 0$
\textcolor{black}{at any doping, whether} for in-plane \textcolor{black}{or} inter-plane transport.
 \textcolor{black}{This shows that there are no nodes in the} superconducting gap. 
However, 
\textcolor{black}{as $x$ decreases into the range of coexistence with antiferromagnetism,
the residual linear term grows more and more rapidly with applied magnetic field.}
\textcolor{black}{This shows that} the superconducting energy gap \textcolor{black}{develops minima at certain locations 
on the Fermi surface and these minima deepen with decreasing $x$.}
%
%
We propose that \textcolor{black}{the minima in the} gap structure \textcolor{black}{arise when} the Fermi surface 
\textcolor{black}{of \K~is reconstructed} by the antiferromagnetic order.

\end{abstract}




\maketitle


\section{Introduction}

Soon after the discovery of superconductivity in iron-based materials, \cite{Hosono}
it was recognized that a conventional phonon-mediated pairing \textcolor{black}{cannot account for}
the high critical temperature \Tc.\cite{phonon}
\textcolor{black}{The observation} of superconductivity in proximity to a magnetic quantum critical point~\cite{Louisreview} 
\textcolor{black}{points instead to} magnetically-mediated pairing,\cite{magneticpairing}
a scenario \textcolor{black}{also} discussed for cuprate and heavy-fermion materials.\cite{Normanscience}
\textcolor{black}{Because such pairing is based on a repulsive interaction, it implies that the superconducting order parameter
must change sign around the Fermi surface.} \cite{Scalapino}
\textcolor{black}{This is the case for the $d$-wave state realized in cuprate superconductors,
where the gap has symmetry-imposed nodes where the Fermi surface crosses the diagonals
at $k_x = k_y$.}
\textcolor{black}{In the $s_\pm$ state proposed for iron-based superconductors,\cite{Mazinspm} 
there are no symmetry-imposed nodes,
but the order parameter has a different sign on the hole and electron pockets.}
\textcolor{black}{One can see that in order to identify the pairing symmetry, associated with a particular pairing mechanism,
it is important to determine the anisotropy of the gap structure.}

\textcolor{black}{In the iron-based superconductors,}
the superconducting gap structure \textcolor{black}{has been studied most extensively in the}
oxygen-free materials with BaFe$_2$As$_2$ (Ba122) as a parent compound.\cite{Rotter}
High-quality single crystals 
\textcolor{black}{can be grown with various} types of dopants to induce superconductivity in the parent \textcolor{black}{antiferromagnet, including:} 
hole doping with potassium in Ba$_{1-x}$K$_x$Fe$_2$As$_2$ (K-Ba122),\cite{Wencrystals}
electron doping with cobalt in Ba(Fe$_{1-x}$Co$_x$)$_2$As$_2$ (Co-Ba122),\cite{Athena,CB}
and iso-electron substitution of arsenic with phosphorus in BaFe$_2$(As$_{1-x}$P$_x$)$_2$ (P-Ba122).\cite{Kasahara}

Early on, an ARPES study of optimally-doped K-Ba122 found a full superconducting gap on all sheets of the Fermi surface.\cite{Ding}
This was explained \textcolor{black}{within the} $s_\pm$ scenario.\cite{MazinNature}
However, \textcolor{black}{subsequent} studies of the superconducting gap structure in Ba122 \textcolor{black}{revealed considerable} diversity. 
In \textcolor{black}{P-Ba122}, the gap is nodal for all \textcolor{black}{dopings}.\cite{HashimotoScience,ShiyanRu}
In \textcolor{black}{Co-Ba122}, the gap is isotropic at optimal doping but it develops \textcolor{black}{nodes} in both under- and over-doped compositions.\cite{GordonPRB,Martin3D,TanatarPRL,Reid3D,Goffryk}
In \textcolor{black}{K-Ba122}, \textcolor{black}{the gap is also} isotropic at optimal doping,\cite{Ding,ReidSUST}
but \textcolor{black}{it develops some $k$-dependence with increasing $x$},\cite{XGLuo,MartinK} and
\textcolor{black}{there are nodes in the gap at $x=1.0$ (KFe$_2$As$_2$)},\cite{Fukazawa,Hashimoto,ShiyanK,ReidPRL,Watanabe,Okazaki} 
\textcolor{black}{where the pairing symmetry may in fact be $d$-wave.\cite{ReidPRL,ReidSUST}}

This diversity in the gap structure \textcolor{black}{has been attributed in part to a}
competition \textcolor{black}{between} intra-band and inter-band pairing interactions.\cite{Chubukovreview,Hirschfeld-ROPP}
\textcolor{black}{Another factor that can affect the gap structure is the presence of a coexisting antiferromagnetic order.\cite{Chubukovreconstruction}}
%
%
%
%
%
%
%
%
%
\textcolor{black}{In this Article,}
we report a study of the superconducting gap structure in \textcolor{black}{K-Ba122} using \textcolor{black}{heat} transport measurements,
\textcolor{black}{for concentrations that cross into the region of the phase diagram where superconductivity and antiferromagnetism coexist.}
\textcolor{black}{We observe that the gap, which is isotropic just above the coexistence region, gradually develops $k$ dependence 
as the magnetic order grows, with minima that deepen with decreasing $x$.
We attribute these minima to the reconstruction of the Fermi surface caused by the antiferromagnetic order.}
%





\begin{figure*} [t]
\centering
\includegraphics[width=18cm]{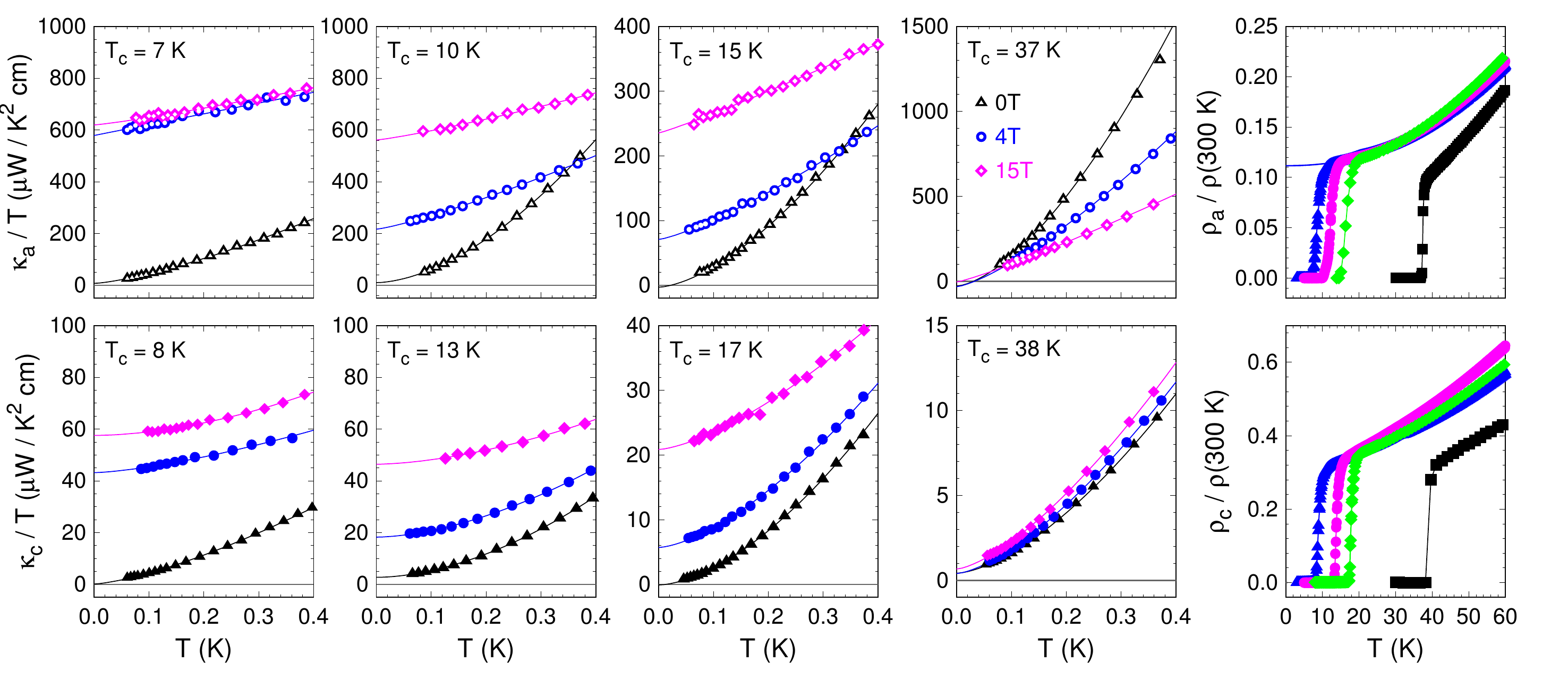}
\caption{
Thermal conductivity of K-Ba122 at four representative K concentrations, 
indicated by their \Tc~values, plotted as $\kappa/T$ vs $T$ for in-plane ($\kappa_ a$, top panels) and inter-plane ($\kappa_c$, bottom panels) directions
\textcolor{black}{of the heat current}. 
Data are shown for three values of the magnetic field, as indicated
(data for other fields are not shown for clarity).
Lines are fits to $\kappa/T = a + bT^{\alpha}$, used to extract the residual linear term 
$a \equiv \kappa_0/T$. 
\textcolor{black}{The} right-most panels show \textcolor{black}{the}~electrical resistivity of the same samples, \textcolor{black}{normalized to its value
at $T = 300$~K,  as a function of temperature}.
\textcolor{black}{
The line is a fit to the data at low temperature, extended to $T=0$ to extract the residual resistivity $\rho_0$.}
}
\label{KTv2}
\end{figure*}

\section{Methods}
Single crystals of \K~were grown using a self-flux technique.\cite{Wencrystals}
\textcolor{black}{Nine samples were cut for $a$-axis transport and seven for $c$-axis transport.
The samples are labelled by their \Tc~value.}
Details of the sample preparation, screening, compositional analysis and resistivity measurements \textcolor{black}{can be found in ref.~\onlinecite{caxis}.}
\textcolor{black}{The technique for making contacts is described in refs.~\onlinecite{SUST} and \onlinecite{patent}.}
The superconducting \Tc~of \textcolor{black}{underdoped} samples changes monotonically with $x$.
\textcolor{black}{We find that the relation between \Tc~and $x$ is well described by the} formula \Tc~$=38.5-54~(0.345-x)-690~(0.345-x)^2$.  
%

The thermal conductivity was measured in a standard one-heater two-thermometer technique \textcolor{black}{described elsewhere},\cite{Reid3D} 
for two directions of the heat flow:
parallel ($Q \parallel c$; $\kappa_c$) and perpendicular ($Q \parallel a$; $\kappa_a$) to the [001] tetragonal $c$ axis. 
The magnetic field $H$ was applied along the $c$ axis. 
Measurements were done on warming after cooling from above \Tc~in a constant field, to ensure a homogeneous field distribution in the sample. 
At least two samples were measured for all compositions to \textcolor{black}{ensure} reproducibility. 
Resistivity measurements \textcolor{black}{to determine} the upper critical field were performed in {\it Quantum Design} PPMS down to 1.8~K.


\section{Results}

\subsection{Electrical resistivity}

\textcolor{black}{In the} right panels \textcolor{black}{of} Fig.~\ref{KTv2},  \textcolor{black}{the resistivity of four K-Ba122 samples, normalized to its value at $T = 300$~K, 
is plotted as a function of temperature,} for both $J \parallel a$ and $J \parallel c$. 
%
\textcolor{black}{The values at 300~K,} $\rho$(300~K), do not \textcolor{black}{change much with doping, and 
$\rho$(300~K) $\simeq 300$ and 1000-2000} $\mu \Omega$ cm, respectively.\cite{caxis,YLiudoping}
%
%
\textcolor{black}{
We use the resistivity curves of each sample to determine \Tc~and the residual resistivity $\rho_0$,
obtained from a smooth extrapolation of $\rho(T)$ to $T = 0$ (see Fig.~\ref{KTv2}).}
\textcolor{black}{We use $\rho_0$ to estimate the residual value of the thermal conductivity in the normal state,}
$\kappa_{\rm N}/T$, via the Wiedemann-Franz law, 
$\kappa_{\rm N}/T = L_0 / \rho_0$,
where $L_0 \equiv (\pi^2/3)(k_{\rm B}/e)^2$.
%

\subsection{Thermal conductivity}
The thermal conductivity \textcolor{black}{of the same four} samples,
\textcolor{black}{measured using the same contacts}, 
is also displayed in Fig.~\ref{KTv2}.
The data in the top row  are for a heat current along \textcolor{black}{the $a$ axis}, 
\textcolor{black}{giving}~$\kappa_a$, \textcolor{black}{while} the data in the bottom panels are for the inter-plane heat current, \textcolor{black}{giving}~$\kappa_c$. 
The fits show that the data below 0.3 K are well described by the power-law function $\kappa/T = a + b T^\alpha$. 
The first term, $a \equiv \kappa_0 /T$, is the residual linear term, entirely due to electronic excitations.\cite{Shakeripour2009a}
The second term is due to phonons, which at low temperature are scattered by the sample boundaries, with $1 < \alpha < 2$.\cite{Sutherland2003,Li2008}
\textcolor{black}{We see that for $H = 0$, $\kappa_{0} /T = 0$ for all samples, within error bars.}
\textcolor{black}{At the highest doping (\Tc~$=37-38$~K), \Kzero~remains negligible even when a magnetic field of 15~T is applied.}
\textcolor{black}{At lower K concentration, however, \Kzero~increases significantly
upon application of a magnetic field.}
\textcolor{black}{Our current data are consistent with our previous measurement of $\kappa_{\rm a}$ in a K-Ba122 sample} with \Tc~$=26$~K.\cite{XGLuo}

\begin{figure} [t]
\centering
\includegraphics[width=8.15cm]{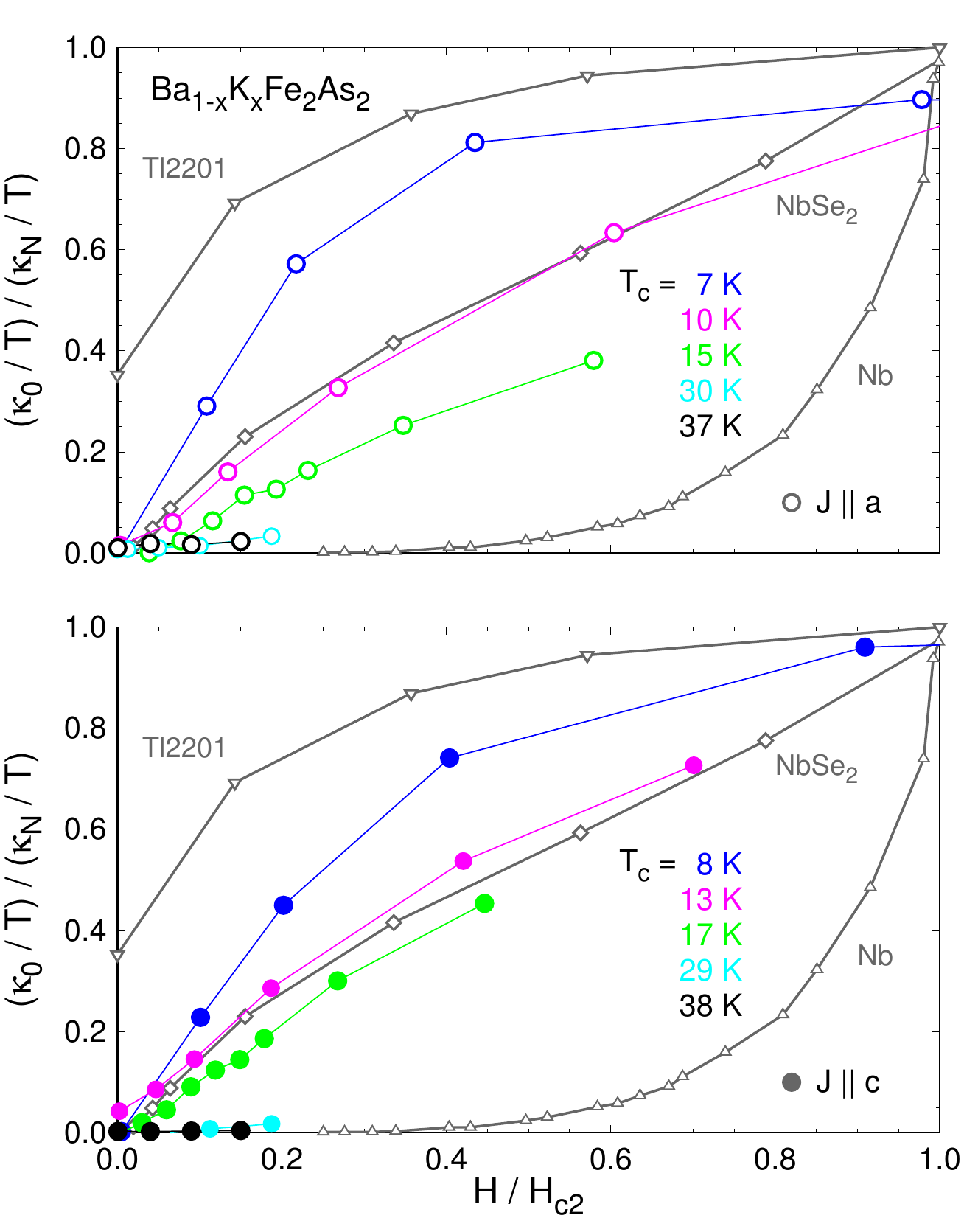}
\caption{
Residual linear term $\kappa_0/T$, normalized by the normal-state conductivity, $\kappa_{\rm N}/T$, as a function of  \textcolor{black}{magnetic field $H$,
normalized by the upper critical field \Hc~(defined in Fig.~3).}
The data for in-plane ($J \parallel a$, open symbols) and inter-plane ($J \parallel c$, closed symbols) transport are shown for five K concentrations, 
indicated by their \Tc~values.
For comparison, we reproduce corresponding data for the isotropic $s$-wave superconductor Nb,\cite{Shakeripour2009a}, 
the multi-band $s$-wave superconductor NbSe$_2$,\cite{Boaknin2003}
and the nodal $d$-wave superconductor Tl-2201.\cite{Proust2002}
}
\label{KHKn}
\end{figure}

\textcolor{black}{Fig.~\ref{KHKn} shows} how the residual linear term $\kappa_0/T$ evolves as a function of magnetic field $H$, 
for both in-plane (top panel) and inter-plane (bottom panel) heat current directions.
\textcolor{black}{In this Figure, $\kappa_{0}/T$ is normalized by the sample's normal-state} conductivity, $\kappa_{\rm N}/T$, 
and the magnetic field is \textcolor{black}{normalized by the sample's} upper critical field  $H_{\rm c2}$ \textcolor{black}{(determined as described in the next section)}. 
%
%
\textcolor{black}{In Fig.~\ref{PDKKn},} the \textcolor{black}{ratio} $\left(\kappa_{0}/T\right) / \left(\kappa_{\rm N}/T \right)$, \textcolor{black}{labelled} 
$\kappa_{0} / \kappa_{\rm N}$ for \textcolor{black}{convenience}, is plotted as a function of K concentration $x$, \textcolor{black}{for two values of the magnetic field:}
$H = 0$ and $H = 0.15$~\Hc. 



\begin{figure} [t]
\centering
\includegraphics[width=9cm]{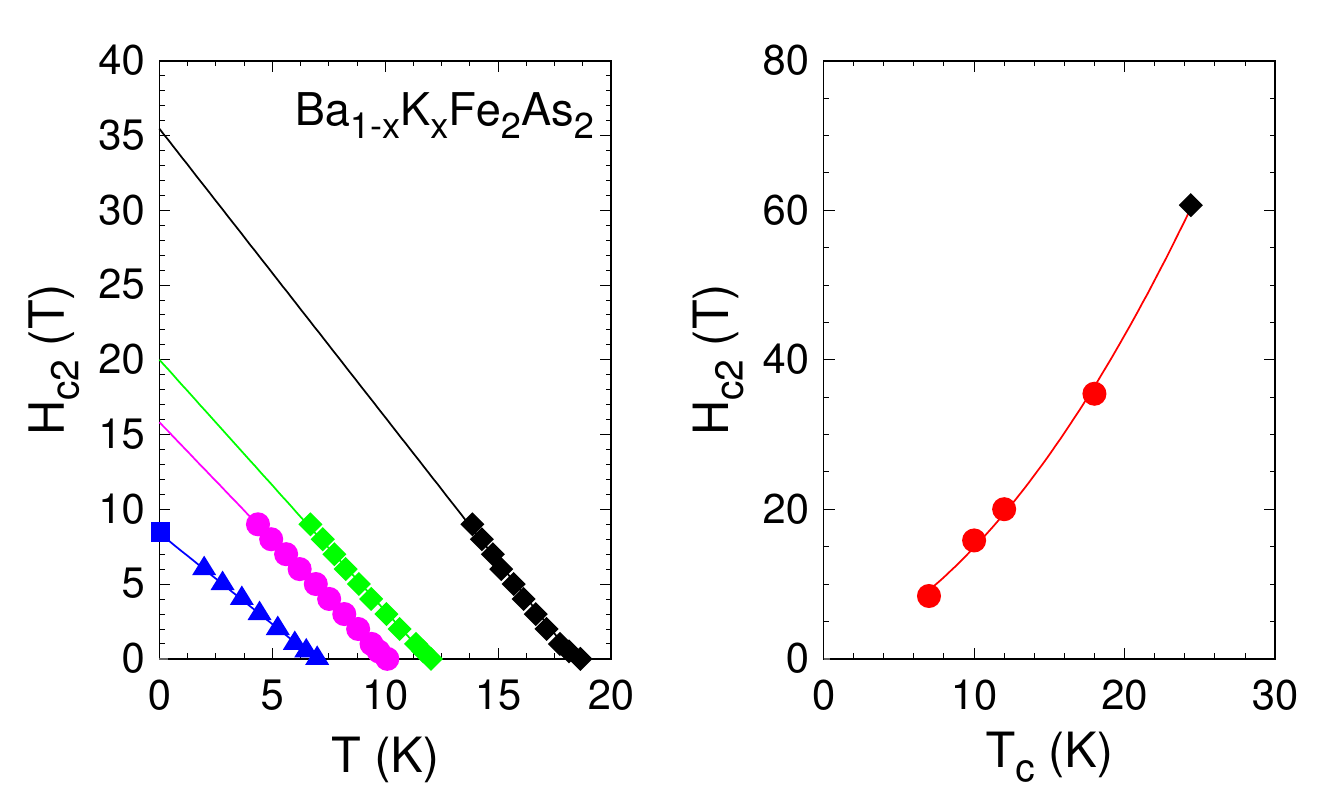}
\caption{
Left panel: 
\textcolor{black}{Upper critical field \Hc$(T)$ for $H \parallel c$, as a function of temperature, for four samples, with \Tc~$=7$~(black), 10 (red), 12 (blue)
and 19~K (green).}
\Hc~is determined from electrical resistivity measurements.\cite{YLiudoping}
The black square shows the \Hc~value determined from thermal conductivity measurements on the \Tc~$=7$~K sample.
\textcolor{black}{
The lines are a linear fit of each data set, extrapolated to $T=0$.
The value of \Hc~thus extrapolated to $T=0$, \Hc(0)~$\equiv$~\Hc, is plotted vs $x$ in the right panel.}
Right panel:
\textcolor{black}{
\Hc(0)~vs \Tc~from data in the left panel (red dots), plus one point at high doping from published data (black diamond).}
\cite{YLiudoping}
%
\textcolor{black}{The line is a polynomial fit of the data, used to extract the values of $H_{c2}$ for the samples with $T_c\in \left[ 7,25 \right]$~K. }
}
\label{Hc2evolution}
\end{figure}

\subsection{Upper critical field}

In the left panel of Fig.~\ref{Hc2evolution}, we plot the \textcolor{black}{upper critical field \Hc~as a function of temperature},
for four \textcolor{black}{values of $x$}, as determined from resistivity measurements for $H \parallel c$. 
For the \textcolor{black}{sample at $x=0.16$ (\Tc~$=7$~K), a field of 15~T} is sufficient to reach the normal state. 
\textcolor{black}{Using the fact that \Kzero~saturates above $H \simeq 9$~T in that sample, we estimate that
\Hc~$= 9$~T at $T \to 0$.
We see that this value agrees with a linear extrapolation of the resistively-determined \Hc$(T)$.}
For the other dopings, we obtain \Hc(0), the value of \Hc$(T)$ at $T \to 0$, by linear extrapolation.
\textcolor{black}{Note that the} slope of the \Hc$(T)$ curves increases with increasing \Tc, 
as expected for superconductors in the clean limit, \cite{YLiudoping}
which holds for K-Ba122 at all dopings.
%
%
\textcolor{black}{In the right panel of Fig.~\ref{Hc2evolution}, we plot \Hc(0) vs \Tc,
including published} data from a sample with a slightly higher concentration.\cite{YLiudoping}



%



\section{Discussion}


\textcolor{black}{From Fig.~\ref{KHKn},
three main characteristics of the gap structure of K-Ba122 can be deduced.
First, the fact that in zero field \Kzero~$=0$ at all dopings, for both current directions, 
immediately implies that there are no zero-energy quasiparticles at $H=0$. 
From this we can infer that there are no nodes 
in the superconducting gap anywhere on the Fermi surface.}

\textcolor{black}{Secondly, we see that the rate at which a magnetic field excites heat-carrying quasiparticles in K-Ba122
varies enormously with doping.}
In the absence of nodes, quasiparticle conduction proceeds by tunnelling between states localized in the cores of adjacent vortices,
which grows exponentially as the inter-vortex separation decreases with increasing field,\cite{Boaknin2003}
\textcolor{black}{as observed in a superconductor with an isotropic gap like Nb (see Fig.~\ref{KHKn}).
The exponential rate is controlled by the coherence length, which is inversely proportional to the gap magnitude.
If the gap is large everywhere on the Fermi surface, the coherence length \textcolor{black}{will}
be small everywhere,
and the growth of $\kappa_0$/\KN~vs $H$ will be very slow at low $H/$\Hc.
This is what we observe in K-Ba122 near optimal doping (\Tc~$\simeq 37$~K).}

\begin{figure}[h]
\centering
\includegraphics[width=8.5cm]{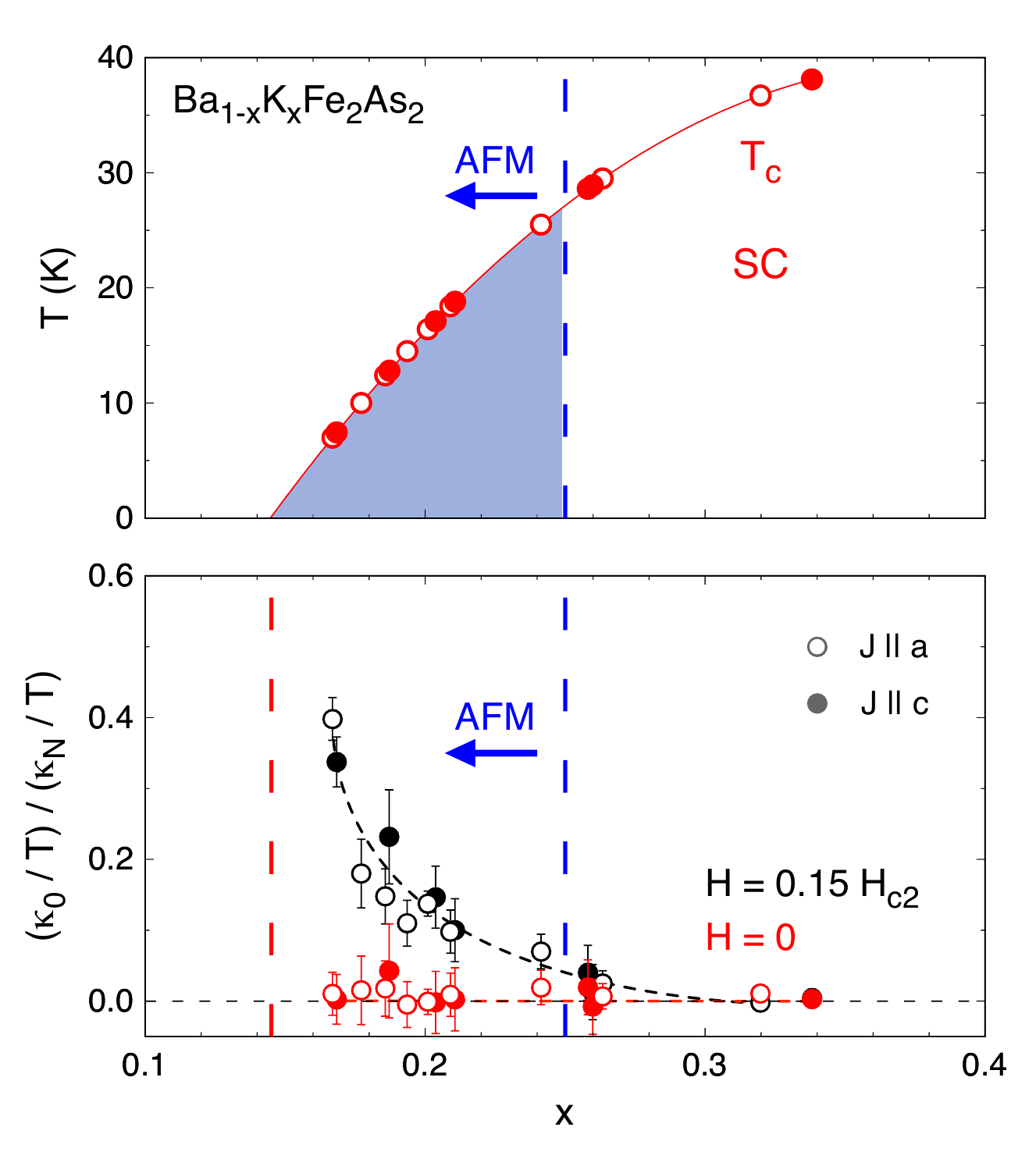}
\caption{
Top panel: 
Doping phase diagram of K-Ba122, showing the onset of the superconducting phase (SC) below the critical temperature 
\Tc~as a function of the K concentration (doping) $x$.
Open (closed) red circles give the \Tc~values of the $a$-axis ($c$-axis) samples used in this study.
For compositions to the left of the dashed blue line at $x \simeq 0.25$ (\Tc~$\simeq 26$~K),\cite{caxis,Avci}
superconductivity coexists with antiferromagnetism (AFM). 
Bottom: 
Residual linear term in the thermal conductivity $\kappa$ as $T \to 0$, $\kappa_0/T$, plotted as a fraction of the normal-state conductivity, 
$\kappa_{\rm N}/T$, 
for both $\kappa_a$ (open symbols) and $\kappa_c$ (closed symbols), 
for magnetic fields $H=0$ (red) and $H=0.15~H_{c2}$ (black). 
Error bars reflect the combined uncertainties of extrapolating $\kappa/T$ and $\rho$ to $T=0$, to get $\kappa_0/T$ and $\rho_0$.
The red vertical dashed line at \textcolor{black}{$x = 0.15$} marks the end of the superconducting phase.
}
\label{PDKKn} 
\end{figure}

\textcolor{black}{If the gap is small on some part of the Fermi surface, compared to the maximum value that dictates \Hc,
this will make it easier to excite quasiparticles, and so lead to an enhanced thermal conductivity at a given value of 
$H /$\Hc.
This is what happens in K-Ba122 with decreasing $x$, whereby $\kappa_0$/\KN~becomes larger and larger with underdoping.
A good way to visualize this evolution is to plot $\kappa_0$/\KN~vs $x$ at $H /$\Hc~$= 0.15$, as done in Fig.~\ref{PDKKn}.
We infer from our in-field data that the gap structure of K-Ba122 develops a minimum somewhere on its Fermi surface,
which gets deeper and deeper with decreasing $x$.}

\textcolor{black}{
There are two ways in which the gap can be small on part of the Fermi surface.
It can develop a strong anisotropy on one sheet of the Fermi surface, as is believed to happen in}
borocarbide superconductors.\cite{borocarbide}
\textcolor{black}{
It can also be small on one surface and large on another.
This multi-band scenario is what happens in MgB$_2$ (ref.~\onlinecite{Sologubenko2002}) and} 
NbSe$_2$ (ref.~\onlinecite{Boaknin2003}).
\textcolor{black}{
In both cases, $\kappa_0$/\KN~grows fast, according to a field scale $H^\star$ much smaller
than \Hc, as it is controlled by the minimum gap.}
See Fig.~\ref{KHKn} for the data on NbSe$_2$.

\textcolor{black}{The third property of the $\kappa_0$/\KN~vs $H$ data in Fig.~\ref{KHKn} is its isotropy with respect to
current direction. The same behaviour is observed for in-plane and inter-plane heat currents.
This implies that the minima which develop in the gap structure have no strong $k_z$ dependence, 
{\it i.e.} they run vertically along the $c$ axis.}


\begin{figure} [t]
\centering
\includegraphics[height=3.5cm]{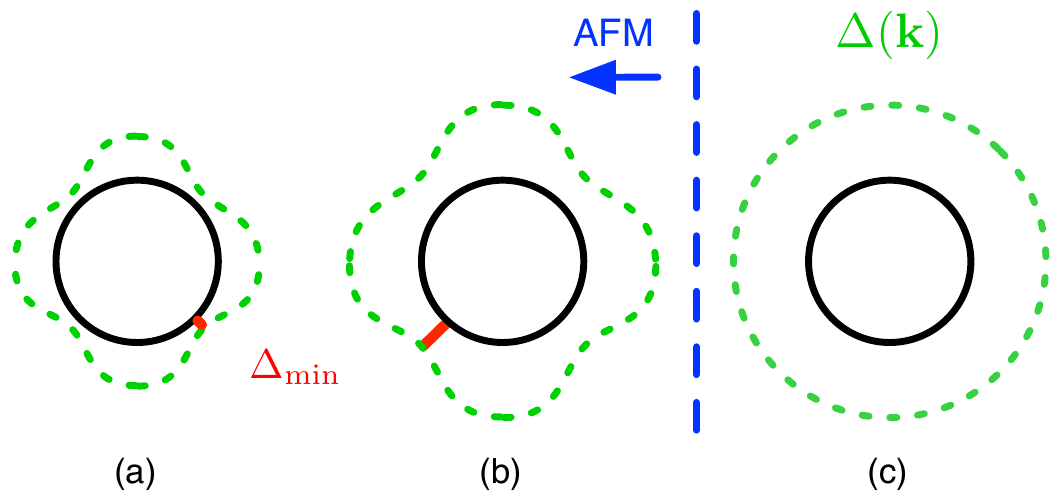}
\caption{
Sketch of the evolution of the gap structure in K-Ba122 with doping $x$, \textcolor{black}{where the gap (green dash line) is shown around a single Fermi surface (black circle).} 
The gap is isotropic at optimal doping (c). 
\textcolor{black}{
The modulation} of the gap starts upon entering the \textcolor{black}{region} where antiferromagnetism (AFM)
appears, and coexists with superconductivity (panel b).
\textcolor{black}{
The gap minimum deepens with decreasing $x$ (panel a).}
}
\label{Sketch}
\end{figure}

\textcolor{black}{
In Fig.~\ref{Sketch}, we provide a  sketch
of how the superconducting gap structure evolves with doping in K-Ba122,
in terms of a simple one-band Fermi surface. 
Since the gap modulation has no significant $k_z$ dependence, we limit our discussion to a 2D picture.
}
At high $x$, the gap is isotropic (panel c), meaning that there is no indication of any modulation of the gap with angle.
Upon lowering $x$, the \textcolor{black}{gap} acquires a \textcolor{black}{modulation}, 
with a minimum gap $\Delta_{\min}$ along some direction \textcolor{black}{(panel b)},
\textcolor{black}{and the gap minimum deepens with decreasing $x$ (panel a).}
This explains 
\textcolor{black}{why the initial rise in $\kappa_0$/\KN~vs $H$ is gets steeper with decreasing $x$ (Fig.~\ref{KHKn}).} 

\textcolor{black}{
The question is: why does the gap develop a modulation?}
In a number of calculations applied to pnictides, the so-called $s_\pm$ state is the most stable.
This is a state with $s$-wave symmetry but with a gap that changes sign in going from the hole-like Fermi surface centred at $\Gamma$ ($\Delta_h > 0 $) to the electron-like Fermi surfaces centred at $X$ and $M$ ($\Delta_e < 0 $).\cite{Wang2009,Graser2009,Chubukovanisotropy}
Although fundamentally nodeless, the associated gap function \textcolor{black}{can have} strong modulations, 
depending on details of the Fermi surface and the interactions, possibly leading to accidental nodes.
\cite{Hirschfeld-ROPP}
The gap modulation comes from a strongly anisotropic pairing interaction, which is also band-dependent, involving \textcolor{black}{the} interplay of intra-band and inter-band interactions. 
It is typically the gap on the electron Fermi surface centred at the $M$ point of the Brillouin zone \textcolor{black}{that} shows a strong angular dependence within the basal plane. \cite{Graser2009,Wang2009}
Therefore, the evolution of the gap structure detected here in K-Ba122, \textcolor{black}{going} from isotropic to modulated \textcolor{black}{with decreasing $x$}, 
is compatible with the general findings of such calculations.
\textcolor{black}{
In Co-Ba122, the development of gap modulations with overdoping was attributed to such a change in interactions.\cite{Reid3D}}

\textcolor{black}{
We propose that another mechanism is at play on the underdoped side of the phase diagram,
having to do with the onset of antiferromagnetic order.}
This is based on the fact that the modulation of the gap and the magnetic order appear at the same concentration, as seen in Fig.~\ref{PDKKn}.

Neutron scattering studies show that antiferromagnetic order in K-Ba122 coexists with superconductivity over a broad range of doping,
up to $x\simeq 0.25$ (\Tc~$\simeq 26$~K), and both magnetism and superconductivity are bulk and occupy at least 95\% of the sample volume.\cite{Avci}
(The fact that $\kappa_0/T = 0$ for $H=0$ in all our samples 
\textcolor{black}{
rules out a scenario of phase separation, whereby significant 
portions of the sample are not superconducting.)} 
%
%
This bulk coexistence is deemed 
\textcolor{black}{to be a strong argument in favour of the $s_\pm$ model, 
and one against the usual $s$-wave scenario.}\cite{Parker,Fernandesspm}

\begin{figure} [t]
\centering
\includegraphics[height=3.5cm]{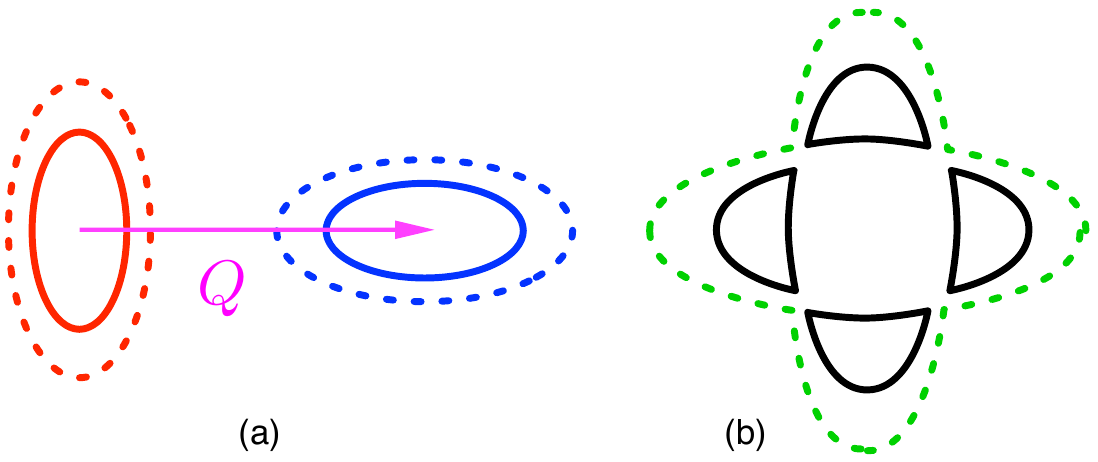}
\caption{
(a) 
Sketch of the evolution of the superconducting gap structure (dashed line) 
in K-Ba122 as the Fermi surface (solid line) is reconstructed
by antiferromagnetic order with a wave-vector $\bf Q$ as drawn.\cite{Chubukovreconstruction}
(b) 
When the hole (red) and electron (blue) pockets overlap as a result of Fermi-surface reconstruction,
an energy gap opens at the crossing points, and this leads to the formation of small crescent-like pieces. 
Calculations show that this can lead to the development of minima in the superconducting gap,
or even nodes.\cite{Chubukovreconstruction}
}
\label{Sketchfolding}
\end{figure}

\textcolor{black}{
Antiferromagnetism in K-Ba122 causes a reconstruction of the Fermi surface
whereby the $\Gamma$-centered hole pocket becomes superpimposed on the edge-centered electron pocket,
as sketched in Fig.~\ref{Sketchfolding}.
Energy gaps open at the points where the original two Fermi pockets cross, 
resulting in the formation of four small crescent-shaped pieces (Fig.~\ref{Sketchfolding}).}
Maiti {\it et al.} \cite{Chubukovreconstruction}
showed theoretically that such a reconstruction triggers a strong modulation 
of the superconducting gap, which develops strong minima, and possibly even
(accidental) nodes, at the crossing points.
It therefore seems natural to attribute the appearance of gap minima 
in underdoped K-Ba122 to the onset of magnetic order.

%
%

%

%

\begin{figure} [t]
\centering
\includegraphics[width=8.3cm]{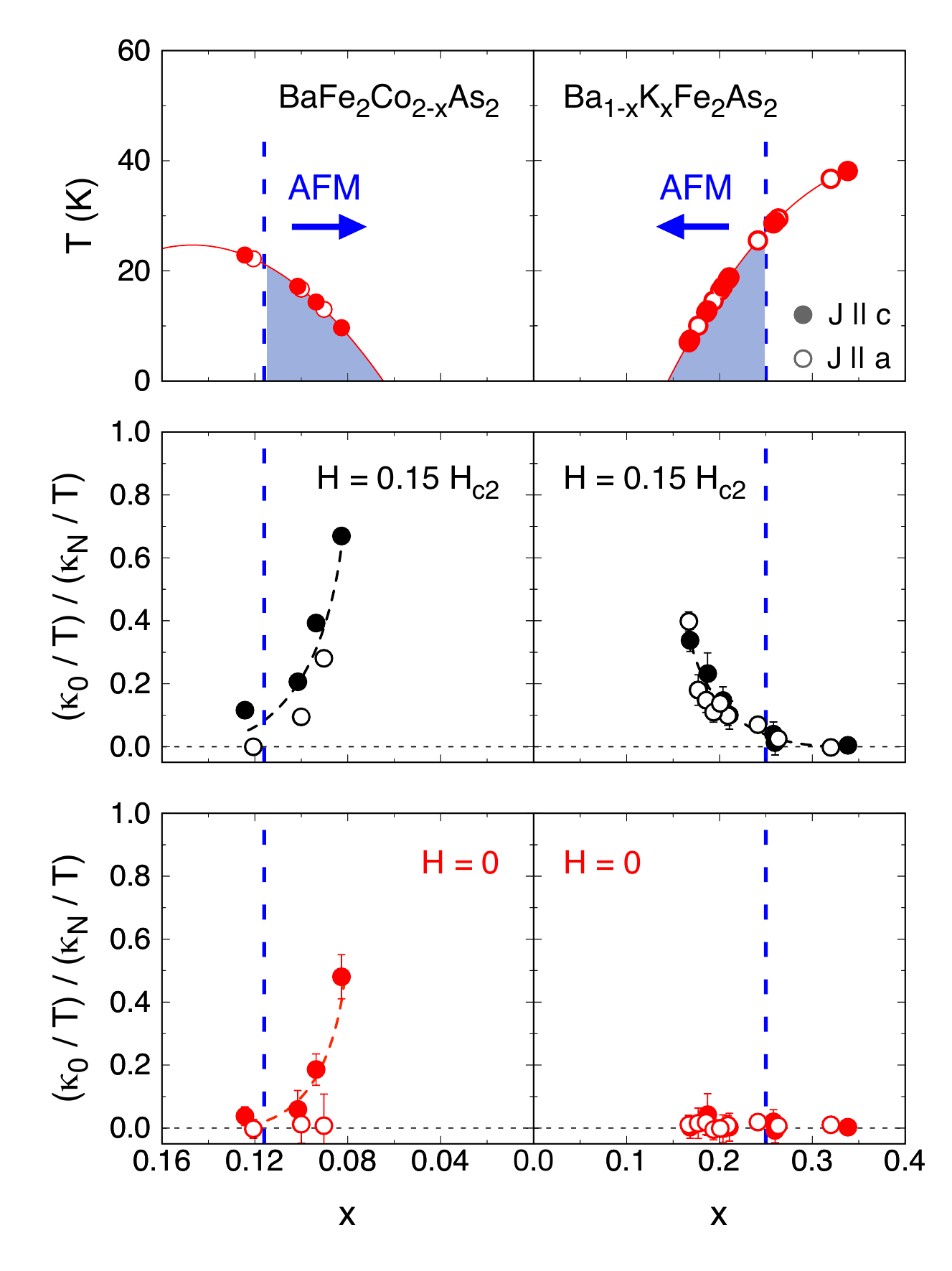}
\caption{
Comparison of the superconducting gap anisotropy 
\textcolor{black}{determined from} thermal conductivity studies in electron-doped Co-Ba122 (left column)\cite{TanatarPRL,Reid3D} 
and hole-doped K-Ba122 (right column, this work). 
The top panels show the phase diagrams of both materials.
\textcolor{black}{The middle} and bottom panels show $\kappa_0/\kappa_{\rm{N}}$ vs $x$ 
for $H=0.15$~\Hc~and $H=0$, respectively. 
Open (closed) symbols correspond to transport along $J \parallel a$ ($J \parallel c$). 
}
\label{Co-K}
\end{figure}

%

\textcolor{black}{
In underdoped Co-Ba122, a similar thermal conductivity study 
revealed that a strong modulation of the gap also appears with the onset of magnetic order.\cite{Reid3D}
So the two materials tell a consistent story.}
In Fig.~\ref{Co-K}, we compare the evolution of the gap as found in thermal conductivity measurements on the two 
sides of the phase diagram of BaFe$_2$As$_2$: the electron-doped side (Co-Ba122) and the hole-doped side (K-Ba122). 
In both cases, the gap is isotropic close to optimal doping, and it develops a strong modulation with underdoping,
\textcolor{black}{concomitant with the onset of antiferromagnetism.}

\textcolor{black}{
However, there is a difference between K-Ba122 and Co-Ba122.
The former does not develop nodes, while the latter does.
These nodes are in regions of the Fermi surface with strong $k_z$ dispersion,
as they give rise to a large zero-field value of \Kzero~for $J || c$, but not for $J || a$ (Fig.~\ref{Co-K}, bottom panel).\cite{Reid3D}
}


\section{Summary}

In summary, the thermal conductivity of K-Ba122 in the $T=0$ limit reveals three main facts.
First, the superconducting gap at optimal doping, where \Tc~is maximal, is isotropic, with no sign of significant modulation anywhere on the Fermi surface.
This reinforces the statement made earlier on the basis of thermal conductivity data in Co-Ba122,~\cite{Reid3D}
that superconductivity in pnictides is strongest when
isotropic, pointing fundamentally to a state with $s$-wave symmetry, at least in the high-\Tc~members of the pnictide family.
Secondly, 
\textcolor{black}{
With underdoping, the superconducting gap becomes small in some parts of the Fermi surface.
The minimum gap gets weaker and weaker with decreasing $x$.
Because this modulation of the gap appears with the onset of antiferromagnetic order, we attribute it
to the associated reconstruction of the Fermi surface.}
Third, 
\textcolor{black}{
although it acquires minima, the superconducting gap structure of underdoped K-Ba122 never develops nodes,
where the gap goes to zero.
This is in contrast with underdoped Co-Ba122, whose gap does have nodes.}

%
%

\section{Acknowledgements}

We thank A.~V.~Chubukov, R.~M. Fernandes, P.~J.~Hirschfeld, D.-H.~Lee and I.~I.~Mazin 
for fruitful discussions and J.~Corbin for his assistance with the experiments. 
\textcolor{black}
{
The work at Sherbrooke was supported by a Canada Research Chair,
the Canadian Institute for Advanced Research, 
the National Science and Engineering Research Council of Canada, 
the Fonds de recherche du Qu\'ebec - Nature et Technologies, 
and the Canada Foundation for Innovation.
The work at the Ames Laboratory was supported by the DOE-Basic Energy Sciences under Contract No. DE-AC02-07CH11358.
The work in China was supported by NSFC and the MOST of China (\#2011CBA00100).
H.S. acknowledges the support of the Iran National Science Foundation.
}


\end{document}